\documentclass[preview,authoryear,3p,onecolumn]{elsarticle}
\usepackage{float,graphicx}
\usepackage{amssymb,amsmath,amsfonts,amstext,bm}
\usepackage{ifpdf,lineno,natbib,url}
\usepackage[colorlinks=true,breaklinks=true]{hyperref}
\usepackage{epstopdf}
\usepackage{graphics}
\usepackage{color}
\usepackage{psfrag}
\usepackage{calc}
\usepackage[english]{algorithm2e}
\usepackage{textcomp}
\usepackage{epstopdf}
\usepackage{accents}
\graphicspath{{./figures/}}
\numberwithin{equation}{section}
\usepackage{lineno}

\usepackage{empheq} 
\usepackage{mathtools} 
\usepackage{color,soul} 
\usepackage{bibentry} 
\usepackage{enumitem}
\usepackage{stmaryrd} 
\usepackage{multirow} 
\usepackage[font=normalsize]{subfig}

\usepackage{color}
\makeatletter
\def\mathcolor#1#{\@mathcolor{#1}}
\def\@mathcolor#1#2#3{%
\protect\leavevmode
\begingroup
\color#1{#2}#3%
\endgroup
}
\makeatother

\usepackage[normalem]{ulem}

\usepackage{etoolbox}
\makeatletter
\def\@mkboth#1#2{}
\newlength\appendixwidth
\preto\appendix{\addtocontents{toc}{\protect\patchl@section}}
\newcommand{\patchl@section}{%
\settowidth{\appendixwidth}{\textbf{Appendix }}%
\addtolength{\appendixwidth}{1.5em}%
\patchcmd{\l@section}{1.5em}{\appendixwidth}{}{\ddt}%
}
\makeatother

\newcommand*\rhoeqs{\rho_{eq}^*}
\newcommand*\oF{\overline{F}}
\newcommand*\onu{\overline{\nu}}
\newcommand*\ok{\overline{\kappa}}
\newcommand*\okp{\overline{\kappa}_p}
\newcommand*\oa{\overline{\alpha}}
\newcommand*\ob{\overline{\beta}}
\newcommand\rz{\overset{0}{\rho}\vphantom{\rho}}

\newcommand\vz{\overset{0}{V}\vphantom{V}}

\newcommand\xz{\overset{0}{x}\vphantom{x}}
\newcommand\cF{\mathcal{F}}

\newcommand\pz{\overset{0}{\mathbf{p}}\vphantom{\mathbf{p}}}
\newcommand{\me}{\mathrm{e}}

\newcommand\rt{\tilde{\rho}}

\newcommand\rtz{\overset{0}{\tilde{\rho}}}

\newcommand{\sn}{\sqrt{\onu}}

\newcommand{\re}{\mathrm{Re}}

\newtheorem{conj}{Conjecture}

\newcommand{\ud}{\, \mathrm{d}}

\begin{document}
\bibliographystyle{elsarticle-harv}

\begin{frontmatter}

\title{Revisiting step instabilities on crystal surfaces. Part~I: The quasistatic approximation}

\author[lms,lpicm,eth]{L.~Guin\corref{ca}}
\ead{laguin@ethz.ch}
\cortext[ca]{Corresponding author}
\author[lms,meca]{M.~E.~Jabbour}
\author[lms,meca,um]{N.~Triantafyllidis}
\bigskip\bigskip

\address[lms]{LMS, \'{E}cole polytechnique, CNRS, Institut Polytechnique de Paris, 91128 Palaiseau, France}
\address[lpicm]{LPICM, \'{E}cole polytechnique, CNRS, Institut Polytechnique de Paris, 91128 Palaiseau, France}
\address[eth]{Mechanics \& Materials Lab, Department of Mechanical and Process Engineering, ETH Zürich, 8092 Zürich, Switzerland}
\address[meca]{D\'{e}partement de M\'{e}canique, \'{E}cole polytechnique, 91128 Palaiseau, France} 
\address[um]{Departments of Aerospace Engineering \& Mechanical Engineering (Emeritus)\\
The University of Michigan, Ann Arbor, MI 48109-2140, USA}

\begin{abstract}
Epitaxial growth on a surface vicinal to a high-symmetry crystallographic plane occurs through the propagation of atomic steps, a process called step-flow growth. In some instances, the steps tend to form close groups (or bunches), a phenomenon termed step bunching, which corresponds to an instability of the equal-spacing step propagation. Over the last fifty years, various mechanisms have been proposed to explain step bunching, the most prominent of which are the inverse Ehrlich-Schwoebel effect (i.e., the asymmetry which favors the attachment of adatoms from the upper terrace), elastically mediated interactions between steps (in heteroepitaxy), step permeability (in electromigration-controlled growth), and the chemical effect (which couples the diﬀusion ﬁelds on all terraces). Beyond the discussion of the influence of each of these mechanisms taken independently on the propensity to bunching, we propose a unified treatment of the effect of these mechanisms on the onset of the bunching instability, which also accounts for their interplay. This is done in the setting of the so-called quasistatic approximation, which by permitting mostly analytical treatment, offers a clear view of the influence on stability of the combined mechanisms.  In particular, we find that the Ehrlich-Schwoebel effect, elastic step-interactions and the chemical effect combine in a quasi-additive fashion, whereas step permeability is neither stabilizing nor destabilizing per se but changes the relative influence of the three aforementioned mechanisms. In a companion paper, we demonstrate and discuss the importance of another mechanism, which we call the dynamics effect, that emerges when relaxing the simplifying but questionable quasistatic approximation.
\end{abstract}


\end{frontmatter}
\newpage
\tableofcontents
\newpage
\section{Introduction}

Thin-film growth gives rise to stresses and surface instabilities, leading to much theoretical work at the interface between mechanics and physics, see, e.g., \citet{Gao1994,Chason2002,Guduru2003,Freund2004} and the references therein. \emph{Epitaxy} is a particular growth technique whereby a crystalline layer is deposited on top of a crystalline substrate. Homoepitaxy refers to the case where both layer and substrate are chemically identical, and heteroepitaxy to the case where the material that makes up the deposited layer is different from that of the substrate. 
Epitaxy is often accompanied by such changes in the surface morphology as the nucleation and evolution of islands \citep{Floro1999,Krug2000}, mound formation, and, when growth occurs on a vicinal surface, as discussed below, the bunching and meandering of atomic steps \citep{Michely2012}.  Understanding the microscopic mechanisms underlying the spontaneous formation of these surface structures is of fundamental interest to crystal growth and of practical interest to the patterning at the nanoscale of semiconductor and metallic surfaces \citep{Tsivion2011,Arora2012}.
 
\begin{figure*}[tb]
\begin{centering}
 \includegraphics[width=0.9\textwidth]{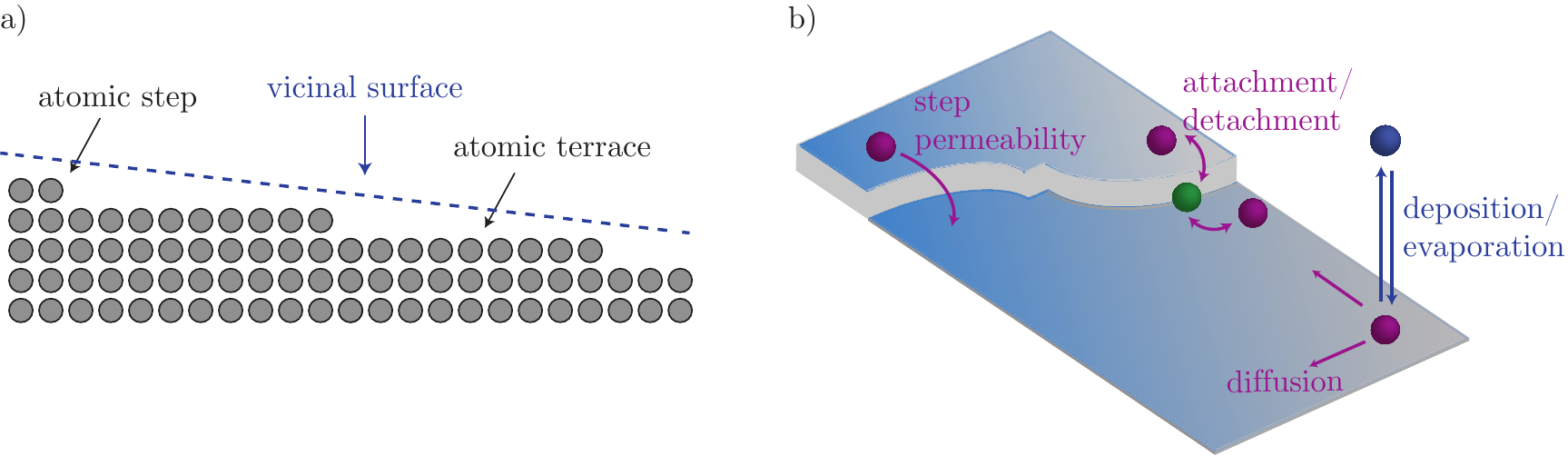}

\caption{a) Schematic of a crystal cut forming a vicinal surface. b)  Schematic of the mechanisms involved in step-flow epitaxial crystal growth. }
\label{fig:setup}
\end{centering}
\end{figure*}

Crystal growth on a \emph{vicinal surface} (i.e., a surface with a slight misorientation relative to a high-symmetry crystallographic plane, see Figure~\ref{fig:setup}(a)) and at low deposition rate occurs through the motion of atomic steps. This step flow involves the diffusion of \emph{adatoms} (i.e., adsorbed atoms) on the terraces that make up the vicinal surface and their attachment to and detachment from steps. Descriptions of the evolution of the surface morphology are at three scales. At the microscopic level, the hopping processes of individual adatoms (or dimers) and the resulting evolution of atomic steps and islands are generally modeled with kinetic Monte Carlo methods, see, e.g. \citet{Chason1990,Myslivecek2002}. At the macroscale, the atomic-scale roughness of the surface---caused by the presence of steps---is neglected and the surface profile is described with a continuous function that evolves both through elastic deformation and by mass rearrangement induced by surface diffusion \citep{Wu1996,Norris1998,Freund1998,Fried2003}. The present work lies at the intermediate \emph{mesoscopic} level, where the diffusing adatoms are accounted for with an adatom density function defined on the successive terraces, which are separated by moving boundaries representing the atomic steps. The corresponding free-boundary problems are referred to as \emph{step-flow models}, the first of which was proposed by \citet{Burton1951} before any direct observation of atomic steps could be made. The advent a few decades later of atomic scale microscopy imaging has fostered many theoretical works on the step-flow model \citep{Krug2005,Misbah2010}. In the mechanics literature, adatoms and atomic steps have generated interest primarily because of the elastic fields that they generate in the crystal bulk and the resulting interactions between multiple steps, adatoms, and between adatoms and steps \citep{Shilkrot1997,Peralta1998,Kukta2002,Kukta2003a,Kukta2003}.

A step changes configuration as a result of the attachment or detachment of adatoms that reach it by diffusion on the adjacent terraces and are supplied via precursors in a vapor, as in chemical vapor deposition, or from heated solid sources in an ultra-high vacuum environment, as in molecular beam epitaxy. At sufficiently low temperatures, the desorption of adatoms is negligible; this deposition regime is the first of two limit cases we consider in this study. In some experiments, the sources are turned off and the substrate on top of which the thin film was deposited is heated radiatively, resulting in the desorption of adatoms from the vicinal surface; this sublimation\footnote{The term \emph{sublimation}, i.e., passage from solid to vapor state is hereby used interchangeably with the term \emph{evaporation}, traditionally used for the passage from liquid to vapor phase.} regime is the second case. A schematic of the elementary processes underlying step-flow growth are shown in Fig.~\ref{fig:setup}(b).

In both regimes, instabilities that affect the shape and/or distribution of steps are observed on semiconductor surfaces, such as Si(111)-7\texttimes7 and GaAs(001), as well as on metallic surfaces, such as Cu(1,1,17). These instabilities are of two types: \textit{meandering}, whereby an initially straight step becomes wavy, and \textit{bunching}, whereby a train of initially equidistant steps evolves into regions of high step density (step bunches) separated by wide terraces. In this two-part article, we are interested in the latter instability, for which the one-dimensional modeling is appropriate.

As mentioned above, we work within the framework of the Burton, Cabrera and Franck (BCF) model. Originally proposed by \cite{Burton1951} for steps that act as perfect sinks for neighboring adatoms (so that the adatom density is continuous at steps and equal to its equilibrium value), it was successively extended by various workers to account for: small deviations from local equilibrium at steps during epitaxial growth \citep{Chernov1961}; possible asymmetries in the attachment of adatoms to, and their detachment from, steps by \cite{Schwoebel1969}, as captured by the direct Ehrlich--Schwoebel barrier or its inverse; interactions between steps, mediated by the elastic fields that they generate in the crystal bulk, which are repulsive in the case of homoepitaxy and attractive in that of heteroepitaxy \citep{Muller2004}; step permeability to the hopping of adatoms between adjacent terraces \citep{Ozdemir1992}; and adatom electromigration when the substrate is heated by a direct current \citep{Stoyanov1991}. Although a detailed literature review is beyond the scope of the present work, the interested reader is
referred to the books of \cite{Saito1996,Pimpinelli1998,Michely2012} and the review articles of  \cite{ Jeong1999,Krug2005,Misbah2010}.

The question of the consistency of the BCF models with the laws of thermodynamics has largely been ignored. It turns out to have implications on the stability predictions of the BCF theory, as this paper and its sequel aim to establish. In particular, the derivation of a thermocompatible BCF-type model by \cite{Cermelli2005} reveals that the configurational force driving the migration of steps, defined as the work-conjugate of the step velocity, has a contribution from the adjacent terraces in the form of the jump of the adatom grand canonical potential. Since the vanishing of this jump is a condition of chemical equilibrium, it should be no surprise that it would contribute to the driving force in an out-of-equilibrim chemical process such as step motion by adatom attachment and detachment. Nonetheless, this contribution, which we refer to as the \textit{chemical effect}, has remained largely unaccounted for in the literature on step flow, and its possible role in the onset of the bunching instability remains mostly unexplored. It is one of the objectives of this two-part article to investigate this role, not just for the step-pairing instability as was previously done in \cite{Cermelli2007}, but for linear perturbations of all wavelengths \citep[see also][]{Cermelli2010}.

There are two facets to the study of the stability of straight steps: the conditions for the onset of the bunching instability and the long-term evolution of step bunches. The present work focuses on the former whereas the latter is addressed elsewhere \citep{Guin2020a,Benoit2020}. Linear-stability investigations of step flow are carried out in the setting of the quasistatic approximation in Part~I, in which certain terms in the moving-boundary problem, namely, the transient term in the reaction-diffusion equation that governs the adatom density on terraces and its advective counterparts in the associated boundary conditions at steps, are neglected. In the few instances where a justification is provided, it is claimed that the quasistatic approximation holds for slow deposition or evaporation \citep{Krug2005,Michely2012}. In Part~II, 
we show that this claim is not well founded: the neglected terms have an impact on the stability of steps with respect to bunching, even in the limit of vanishingly small deposition and evaporation rates. They give rise collectively to an additional stabilizing/destabilizing mechanism which we refer to as the \textit{dynamics effect}.  Furthermore, while the neglect of the dynamics terms is justified when computing the fundamental solution corresponding to a train of equidistant steps in the regime of low deposition or evaporation, there is no basis for ignoring them in the system that governs small perturbations about the aforementioned fundamental solution. Hence, the stability results under the quasistatic approximation ought to be considered with caution. While there may be conditions under which the influence of the dynamics effect on stability is negligible in comparison with that of other mechanisms (e.g., elastic interactions), rendering valid the predictions of the quasistatic linear-stability analysis, we have no criteria by which to determine \emph{a priori} that such conditions are indeed satisfied.

Nevertheless, the quasistatic approximation is an important mathematical simplification which, because it allows a mostly analytical treatment of the linear-stability problem, affords physical insight into the stabilizing or destabilizing influence of each of the basic mechanisms underlying step flow, including the chemical effect, and how their interplay controls the onset of bunching. For these reasons, in Part~I of this work, we adhere provisionally to the quasistatic approximation, aware that in doing so, we miss one stabilizing/destabilizing mechanism: the dynamics effect. The latter will be addressed in Part~II, where we relax the quasistatic approximation through the use of a more involved stability analysis.

The rest or the article is organized as follows:  in Section~\ref{sec:model}, we introduce the equations governing the step-flow problem. The linear
stability analysis corresponding to bunching is developed in Section~\ref{sec:stability1}. We present the results in Section~\ref{sub2:mech} and discuss them 
in Section~\ref{sec:conclusion}.

\section{Problem formulation}
\label{sec:model}

We provide in Section~\ref{sub:goveq} a quick overview of the equations that govern, in the form of a moving-boundary problem, step-flow growth and sublimation. In Section~\ref{sub:thermo}, we briefly discuss the origin of the step boundary conditions with an emphasis on the contribution to the driving force at each step of the adatoms on its adjacent terraces, one unaccounted for in the various extensions of the BCF model and to which we refer as the \emph{chemical effect}. This is followed in Section~\ref{sub:elas} by a short review of another contribution to the driving force acting on a given step, namely, that of the elastic fields generated in the bulk of the crystal by the remaining steps on its free surface; departing from a common assumption in the literature on step-flow epitaxy, we do not restrict this contribution to nearest-neighbor interactions. In Section~\ref{sub:nd}, we nondimensionalize the moving-boundary problem of Section~\ref{sub:goveq}, which yields several dimensionless numbers that quantify the relative strengths of the competing kinetic and energetic mechanisms underlying step dynamics. Finally, the quasistatic approximation is presented in 
Section~\ref{sub:approx}.

\subsection{Moving-boundary problem} 
\label{sub:goveq}

\begin{figure*}[tb]
\begin{center}
\includegraphics[width=80mm]{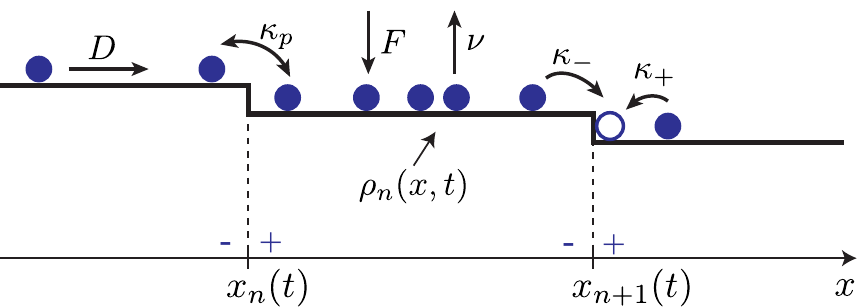}
\caption{\label{fig:step} Schematic of two successive steps located, at time $t$, at $x_n(t)$ and $x_{n+1}(t)$. The atomic mechanisms involved in step flow are: adatom diffusion on terraces, with diffusion coefficient $D$; adsorption from, and desorption to, a vapor or an ultra-high vacuum environment, with deposition rate $F$ and evaporation coefficient $\nu$; attachment of adatoms to, and their detachment from, steps, with kinetic coefficients $\kappa_+$ and $\kappa_-$ as each step is approached from its lower and upper adjacent terraces, respectively; adatom hopping across steps, with step permeability $\kappa_p$.}
\end{center}
\end{figure*}

We consider an infinite sequence of straight steps, so that ours is a one-dimensional setting in which $x_n(t)$ denotes the position of the $n$th step ($n\in\mathbb{Z}$) at time $t$. Each step migrates as a result of the attachment or detachment of adatoms that diffuse on its adjacent upper and lower terraces. These terrace adatoms are supplied from a vapor phase, either through chemical vapor deposition or in an ultra-high vacuum environment through molecular beam epitaxy. The schematics of this problem and the associated mechanisms are shown in Fig.~\ref{fig:step}.

Our objective is to analyze the influence on the stability of steps against bunching of various physical processes that have been added over the last decades to the original step-flow model of Burton, Cabrera, and  Frank \citep{Burton1951}. These consist of: (i) the Ehrlich--Schwoebel barrier \citep{Schwoebel1966,Schwoebel1969} or its inverse, which embodies a possible asymmetry in the kinetics associated with the attachment of terrace adatoms to, or their detachment from, steps; (ii) elastically mediated step-step interactions, present during both homo- and hetero-epitaxy \citep{Tersoff1995}, which derive from the contribution to the driving force acting on each step of the elastic fields generated by the remaining steps that make up the vicinal surface \citep{marchenko1980,stewart1994}; (iii) the permeability of steps or lack thereof, which allows or prevents the direct hopping of adatoms between adjacent terraces \citep{Ozdemir1992,Pierre-Louis2003}; and, finally, (iv) the coupling at each step between the diffusion fields on its adjacent terraces, which derives from the energetic contribution to the configurational force at the step of nearby adatoms \citep{Cermelli2005,Cermelli2007}, and to which we refer as the chemical effect. Note that, in the above list of physical ingredients whose role in the bunching of steps we wish to investigate, we have not included adatom electromigration on terraces \citep{Latyshev1989,Stoyanov1991,Yang1996,Fu1997,Degawa2000,Stoyanov2000,Degawa2001,Zhao2004}. In doing so, we have deliberately restricted the scope of the present study to experiments in which the substrate on top of which epitaxial growth or sublimation occurs is heated radiatively, as opposed to heating by an electric current. Indeed, surface electromigration adds another layer of complexity to the study of step instabilities, namely the multiple stability reversals that are observed as the temperature at which growth or sublimation occurs is varied, and is addressed elsewhere \citep{Benoit2020b}.

We begin with a statement of the moving-boundary problem that governs step flow, whose unknowns are the terrace adatom densities $\{\rho_n(x,t)\}_{n\in\mathbb{Z}}$ and step positions $\{x_n(t)\}_{n\in\mathbb{Z}}$. The reaction-diffusion equation and associated step conditions that make up this problem are derived elsewhere \citep{Cermelli2005} in the absence of step permeability and neglecting elastic interactions between steps, two mechanisms that are accounted for in the present study and briefly discussed in Sections~\ref{sub:thermo} and~\ref{sub:elas}. This moving-boundary problem is an approximation of a more general one \citep[see][]{Cermelli2005,Guin2018,Benoit2020b} valid in the limit of small departures of the adatom density from its step equilibrium value $\rho_{eq}^*$, i.e., whenever 
\begin{equation} \label{eq:linear}
\vert\rho_n(x,t)-\rho_{eq}^*\vert\ll\rho_{eq}^*.
\end{equation}
for all $x \in (x_n(t),x_{n+1}(t))$ and all $n\in\mathbb{Z}$.
The assumption \eqref{eq:linear}, underlying the step-flow problem presented below, is omnipresent in the literature on step instabilities, although seldom made explicit. Its conditions of validity can be specified in terms of the physical parameters of the crystal growth by computing the adatom density of the steady-state solution to step flow, as we do in Section~\ref{sub:sss}.

Let $D$, $F$, and $\nu$ be the adatom diffusivity, deposition flux, and desorption coefficient, respectively. Assuming that adatoms behave like an ideal lattice gas, mass balance on the terrace $(x_n(t),x_{n+1}(t))$ gives
\begin{equation} \label{eq:diff1}
\partial_t \rho_n=D \partial_{xx} \rho_n+F- \nu \rho_n.
\end{equation}

Next, denote by $\kappa_+$ and $\kappa_-$ the kinetic coefficients for the attachment of adatoms to, and their detachment from, a step as they approach it from the lower and upper adjacent terraces, respectively, and let $\kappa_p$ be the permeability coefficient associated with the hopping of adatoms between adjacent terraces.
Writing $J^+_n$ for the adatom current into the $n$th step from its lower adjacent terrace and $J^-_n$ for its counterpart from the upper terrace, the reaction-diffusion equation \eqref{eq:diff1} is supplemented by boundary conditions that derive from the localization of mass balance as the $(n+1)$th step is approached from the left and the $n$th step from the right (Fig.~\ref{fig:step}), respectively,
\begin{align}\label{eq:bc1}
\begin{aligned}
-\rho_n^- \dot{x}_{n+1} - D (\partial_x\rho_n)^-&=\underbrace{\kappa_-\left[\rho_n^- - \rho_{eq}^* 
- a^2 \rho_{eq}^* \left(\chi \llbracket \rho \rrbracket_{x_{n+1}} - \frac{\mathfrak{f}_{n+1}}{k_{_B} T}
- \frac{\dot{x}_{n+1}}{k_{_B} T b} \right)\right]}_{J_{n+1}^-}  - \kappa_p \llbracket \rho \rrbracket_{x_{n+1}},\\[4pt]
\rho_n^+ \dot{x}_{n} + D (\partial_x\rho_n)^+&=\underbrace{\kappa_+\left[ \rho_n^+ - \rho_{eq}^* 
- a^2 \rho_{eq}^* \left(\chi \llbracket \rho \rrbracket_{x_{n}} - \frac{\mathfrak{f}_{n}}{k_{_B} T} 
- \frac{\dot{x}_{n}}{k_{_B} T b}\right)\right]}_{J_{n}^+}- \kappa_p \llbracket \rho \rrbracket_{x_{n}},
\end{aligned}
\end{align}
where $\dot{x}_n(t)$ is the velocity of the $n$th step, $a^2$ the surface area occupied by a crystal atom, $k_{_B}$ the Boltzmann constant, $T$ the absolute temperature, $b$ the step kinetic modulus, $\mathfrak{f}_n(t)$, whose expression is discussed in Section~\ref{sub:elas}, is the elastic contribution of the remaining steps to the configurational force acting on the $n$th step. In \eqref{eq:bc1}, the superscripts $-$ and $+$ denote the limiting values at each step of discontinuous terrace fields as the step is approached from above and below, respectively, and $\llbracket\rho\rrbracket_{x_{n}}:=\rho_{n+1}(x_n(t),t)-\rho_n(x_n(t),t)$ is the jump of the adatom density across the $n$th step. Finally, $\chi$ is a parameter, which we introduce for convenience, with value $0$ or $1$. In fact, as we shall see in Section~\ref{sub:thermo}, the thermodynamics of nonequilibrium processes dictates that $\chi=1$, so that the jumps $-a^2\rho^*_{eq}\llbracket\rho\rrbracket_{x_{n+1}}$ and $-a^2\rho^*_{eq}\llbracket\rho\rrbracket_{x_n}$, which are unaccounted for in the standard BCF model, turn out to be intrinsic to the step conditions \eqref{eq:bc1}. Nonetheless, setting $\chi=0$ affords us insight into the separate influence of the basic mechanisms underlying step flow on the onset of step bunching, since by doing so we can formally eliminate the chemical effect from the linear stability analysis.

Finally, localization of mass balance at the $n$th step yields the interfacial condition
\begin{equation} \label{eq:velocity1}
\dot{x}_n=a^2 (J_{n}^- + J_{n}^+),
\end{equation}
which, as intuitively expected, states that the rate at which the step advances or recedes is proportional to the net flux $J_n^-+J_n^+$ of adatoms from the adjacent terraces. Solving the moving-boundary problem \eqref{eq:diff1}--\eqref{eq:velocity1} for all $n\in \mathbb{Z}$ delivers the adatom distribution on all terraces and the step positions at all times.

\subsection{Step boundary conditions with chemical coupling between terraces 
\label{sub:thermo}}

In this section, we give some insight on the origin and physical interpretation of the different terms entering the boundary conditions \eqref{eq:bc1}. Consider \eqref{eq:bc1}$_2$, which holds at the $n$th step, it differs from the standard condition found in the literature on step dynamics \citep[see, e.g.,][]{Pierre-Louis2003} by two terms: $-a^2\rho^*_{eq}\llbracket\rho\rrbracket_{x_n}$ and $a^2\rho^*_{eq}\dot{x}_n(t)/k_{_B} T b$. The origin of each is to be found in the dissipation inequality, as localized at a generic step. To see this, we briefly review the argument of \cite{Cermelli2005,Cermelli2007}, specializing it on one hand to the present one-dimensional setting and extending it on the other to account for elastically mediated step-step interactions and adatom hopping between adjacent terraces.

For notational simplicity, we omit in what follows all indices that label steps and terraces. Let $V$ be the step velocity, and denote by $J_+$ and $J_-$ the adatom fluxes into, or out of, the step from its adjacent lower and upper terraces. Finally, write $J_p$ for the net flux across the step of adatoms hopping from the upper terrace onto the lower one. As the step is approached from above and below, localization of mass balance yields the conditions
\begin{equation} \label{eq:adbal}
\begin{aligned}
-\rho^- V + \jmath^- &=  J_-  + J_p,\\[4pt]
\rho^+ V - \jmath^+ &=  J_+  - J_p,
\end{aligned}
\end{equation}
with $\jmath$ the adatom diffusive flux on terraces. In \eqref{eq:adbal}$_1$, the left-hand side is the flux of adatoms from, or to, the upper terrace, with its  advective and diffusive components; the right-hand side is the sum of the flux of adatoms that attach to, or detach from, the step and the flux of adatoms that hop onto the lower terrace; mass balance dictating the equality of the two sides. A similar interpretation holds for \eqref{eq:adbal}$_2$.

As growth or sublimation occurs at fixed temperature, we restrict our attention to isothermal settings in which the first and second laws of thermodynamics combine to deliver a free-energy imbalance that serves, as we shall see below, to identify the configurational force that drives step motion and to impose restrictions on the constitutive relations for the fields that appear in \eqref{eq:adbal}. Let $\psi(x,t)=\hat{\psi}(\rho(x,t))$ be the adatom free-energy density (per unit area of the terrace) and $\mu(x,t)=\partial_\rho\hat{\psi}(\rho(x,t))$ the adatom chemical potential, and denote by $\mu_s(t)$ the step chemical potential.\footnote{We neglect surface elasticity, otherwise the adatom free-energy density and chemical potential would depend also on the terrace stretch.} Localization of the free-energy imbalance at the step delivers the interfacial dissipation inequality
\begin{equation} \label{eq:dissip}
\underbrace{\left(\frac{\mu_s}{a^2} + \llbracket \psi - \mu \rho \rrbracket -\psi_c + \mathfrak{f} \right)}_{\digamma}V + (\mu^- - \mu_s ) J_- + (\mu^+ - \mu_s ) J_+ +(\mu^- -\mu^+) J_p \geq 0,
\end{equation}
where $\llbracket \psi - \mu \rho \rrbracket$ is the jump across the step of the adatom grand canonical potential, $\psi_c$ is the areal free-energy density of the \emph{undeformed} crystal (which, in the absence of bulk diffusion, is constant), and $\mathfrak{f}$ is the contribution to the configurational force $\digamma$ driving the step motion of the elastic fields generated by the remaining steps on the vicinal surface, 
see Section~\ref{sub:elas} below. The linear constitutive relations between the (generalized) velocities $V$, $J_-$, $J_+$, and $J_p$ on one hand and their conjugate driving forces $\digamma$, $\mu^--\mu_s$, $\mu^+-\mu_s$, and $\mu^--\mu^+$ on the other, 
\begin{equation} \label{eq:cr1}
\left\{\begin{aligned}
V & =b \left(\frac{\mu_s}{a^2} + \llbracket \psi - \mu \rho \rrbracket -\psi_c + \mathfrak{f}\right), \\[4pt]
J_- & = \gamma_- (\mu^- - \mu_s) \quad \text{and} \quad J_+ = \gamma_+ (\mu^+ - \mu_s),\\[4pt]
J_p &= \gamma_p (\mu^- - \mu^+),
\end{aligned}
\right.
\end{equation}
are sufficient for \eqref{eq:dissip} to hold for any step-flow process, provided that the kinetic modulus $b$, attachment-detachment coefficients $\gamma_+$ and $\gamma_-$, and permeability $\gamma_p$ are nonnegative. By \eqref{eq:dissip} and \eqref{eq:cr1}, the dissipation $\mathcal D$ per unit length of the step is given by
\begin{equation}
{\mathcal D} = \underbrace{b\digamma^2+(\mu^--\mu_s)J_-+(\mu^+-\mu_s)J_+}_{\text{dissipation due to adatom}\atop\text{attachment-detachment}}\underbrace{-\llbracket\mu\rrbracket J_p}_{\text{dissipation due to}\atop\text{adatom hopping}}\geq0.
\end{equation}

The kinetic relation \eqref{eq:cr1}$_1$ can be understood by recalling that equilibium in a two-phase body involves more than one condition at the interface separating them. For instance, phase equilibrium between the solid and liquid phases of a pure substance implies the Gibbs--Thomson relation, while energy balance prescribes the continuity of the heat flux at the solidification front \citep{Davis2001}. Likewise, in the mechanical setting of solid-to-solid phase transformations, phase equilibrium yields the Maxwell relation---which imposes the continuity of the normal component of Eshelby's energy-momentum tensor---while force balance imposes the continuity of traction at the interphase \citep{Abeyaratne2006}. Finally, in transformations driven by the transfer of matter from one phase to the other (e.g., in phase separation problems),  mass balance imposes the continuity of species fluxes and phase equilibrium dictates that the grand canonical potential be continuous at the phase boundary (when the interfacial energy is negligible), which is the chemical analogue of the thermal Gibbs--Thomson relation or the mechanical Maxwell relation \citep{Gurtin1996}. Since crystal growth by step flow is driven by the \emph{out-of-equilibrium chemical process} of attachment of adatoms to, and their detachment from, atomic steps which are endowed with a thermodynamic structure, it is natural that the mass balance
\begin{equation}
V=a^2(J_-+J_+),
\end{equation}
with the adatom fluxes into, or out of, the step given by \eqref{eq:cr1}$_2$, should replace the continuity of the adatom flux, and that the kinetic relation \eqref{eq:cr1}$_1$ should generalize the continuity of the grand canonical potential by linking the step velocity $V$ to the configurational force $\digamma$ driving it. Note that \eqref{eq:cr1}$_1$ can be rewritten as
\begin{equation} \label{eq:mus}
\mu_s=\mu_c-a^2 \left(\mathfrak{f} + \llbracket \psi - \mu \rho \rrbracket  - \frac{V}{b}\right),
\end{equation}
with $\mu_c:=a^2\psi_c$ the areal free-energy density of the underformed crystal. Thus, \eqref{eq:mus}  makes clear that the step chemical potential $\mu_s$ differs from $\mu_c$ and that the difference involves, in addition to the elastic step-step interaction  $\mathfrak{f}$ found in the literature \citep[cf., e.g., ][]{Tersoff1995}, two contributions: one is energetic, in the form of the jump $\llbracket \psi - \mu \rho \rrbracket$ of the adatom grand canonical potential, while the other, $-V/b$, akin to the kinetic undercooling found in solidification problems, is dissipative.

Finally, recalling our assumption that the terrace adatoms behave like an ideal lattice gas, 
\begin{equation} \label{eq:psi}
\hat{\psi}(\rho)=\rho\left\{k_{_B} T \bigg[\ln\bigg(\frac{\rho}{\rhoeqs}\bigg)-1\bigg] + \mu_c \right\},
\end{equation}
from which it follows that $\llbracket \psi - \mu \rho \rrbracket=-k_{_B} T \llbracket \rho \rrbracket$. The linearization of the expression of the adatom chemical potential deriving from \eqref{eq:psi} on differentiation with respect to $\rho$ and substitution into \eqref{eq:cr1}$_{2,3}$ delivers \eqref{eq:bc1} as an approximation of \eqref{eq:adbal}, valid when the deviation of the adatom density from its step equilibrium value remains small, provided that the relations
\begin{equation}
\kappa_\pm:=\frac{k_{_B} T}{\rhoeqs} \gamma_\pm \quad \text{and} \quad \kappa_p:=\frac{k_{_B} T}{\rhoeqs}\gamma_p
\end{equation}
defining the attachment-detachment and permeability coefficients for the linearized model in terms of their counterparts for the nonlinear model are adopted.

\subsection{Elastic interactions between steps} 
\label{sub:elas}

Atomic steps are defects on the free surface of a crystal, each generating a strain field in the bulk.\footnote{We work in the setting of isotropic linear elasticity, so that each of the net elastic (displacement, strain, and stress) fields in the bulk is the sum of the corresponding fields generated by the individual steps.} For sufficiently thick films, the crystalline bulk can be assimilated to a semi-infinite medium, and the elastic field generated by each step is computed by replacing it, in the setting of homoepitaxy, with a force dipole with tangential and normal dipole moments $d_x$ and $d_z$, 
and, in the case of heteroepitaxy, by adding to the force dipole a force monopole of moment $m$ \citep{marchenko1980,stewart1994,Tersoff1995}. Consequently, the $n$th step is subjected to the configurational force $\mathfrak{f}_n$ resulting from the interaction of its elastic field with those of the remaining steps on the vicinal surface. Letting $R \in \mathbb{N}$ be the arbitrary range of step-step elastic interactions, i.e., the range beyond which they become negligible. It can be shown that this elastic contribution to the configurational force acting on the $n$th step is given by
\begin{equation} \label{eq:elasCF}
\mathfrak{f}_n=\sum_{\substack{r\in\{-R,...,R\} \\ r \neq 0 } } \left\{\frac{\beta}{x_{n+r}-x_n} - \frac{\alpha}{(x_{n+r}-x_n)^3}\right\},
\end{equation}
where the coefficients $\alpha>0$ and $\beta \geq 0$ account for the dipole-dipole and monopole-monopole interactions between steps, respectively, and are derived from the dipole and monopole moments through the relations
\begin{equation} \label{eq:alphabeta}
\alpha=\frac{4(1-\nu^2)(d_x^2+d_z^2)}{\pi E} \quad  \mbox{and} \quad \beta=\frac{2(1-\nu^2)m^2}{\pi E},
\end{equation}
with $E$ the Young modulus and $\nu$ the Poisson ratio of the bulk material \citep{marchenko1980,Muller2004,Guin2018}. 

Importantly, \eqref{eq:elasCF} shows that the monopole-monopole interactions are attractive whereas their dipole-dipole counterparts are repulsive. Finally, one can interpret the effect of $\mathfrak{f}_n$ in \eqref{eq:bc1} as changing the equilibrium adatom density from its value $\rho_{eq}^*$  for an isolated step to $\rho_{eq}:=\rho_{eq}^*\left(1-{a^2\mathfrak{f}_n}/{k_{_B} T}\right)$ for a step that interacts elastically with the other steps.

\subsection{Nondimensionalization} 
\label{sub:nd}

Hereafter, we nondimensionalize $x$ by the initial terrace width $L_0$, $t$ by the characteristic time $L_0^2/D$ for adatom diffusion, and $\rho$ by its equilibrium value $\rho^*_{eq}$ at a straight step in the undeformed crystal. In doing so, we introduce the parameters of the dimensionless moving-boundary problem. 

We begin with the equilibrium adatom coverage
\begin{equation}
\Theta:=a^2 \rhoeqs\in(0,1)
\end{equation}
which measures the fraction of occupied lattice sites when the adatom density takes on its equilibrium value. 

Next, we define the dimensionless deposition rate
\begin{equation} \label{eq:oF}
\overline{F}:=\frac{F L_0^2 }{\rhoeqs D}=  \frac{L_0^2}{\big(L_d^{dep}\big)^2}
\end{equation}
as the (square of the) ratio of the intial terrace width to the diffusion length under deposition $L_d^{dep}:=\sqrt{\rho^*_{eq}D/F}$. Similarly, the dimensionless desorption coefficient
\begin{equation}
\overline{\nu}:=\frac{\nu L_0^2}{D}= \frac{ L_0^2 }{\big(L_d^{eva}\big)^2}
\end{equation}
measures the initial terrace width relative to the diffusion length under deposition $L^{eva}_d:=\sqrt{D/\nu}$.\footnote{The requirement $\rho \le a^{-2}$ that the adatom density remain below the density of the underlying crystal and the stricter assumption that it deviates little from its equilibrium value $\rho^*_{eq}$ impose upper bounds on the values that $\overline{F}$ and $\overline{\nu}$ can take, as is discussed in Appendix~C of  Part~II of this work.}

Moreover, given that the attachment and detachment at each step of adatoms from its upper and lower terraces are governed by similar atomistic processes, the associated kinetic rates are expected to have comparable orders of magnitude, as measured by 
\begin{equation} \label{eq:ok}
\ok:=\frac{\kappa_- L_0}{D}
\end{equation}
which quantifies the ratio of the initial terrace width to the kinetic length $D/\kappa_-$ \citep{Krug2005}. We can interpret $\ok$ as the ratio of the velocity $\kappa_-$ which characterizes the attachment and detachment of adatoms at steps to the velocity $D/L_0$ which characterizes their diffusion on terraces. We can therefore distinguish between two limits: $\ok \ll 1$ corresponds to the \emph{attachment/detachment-limited regime} in which adatom attachment/detachment at steps is the rate-limiting kinetic process, whereas $\ok \gg 1$ is associated with the \emph{diffusion-limited regime} in which it is the kinetics of adatom diffusion on terraces that is rate-limiting. 

In the same vein, the weight of adatom hopping between terraces (relative to adatom diffusion) is quantified by the dimensionless permeability coefficient 
\begin{equation} \label{eq:okp}
\okp:=\frac{\kappa_p L_0}{D}
\end{equation}
which measures the hopping velocity $\kappa_p$ relative to its diffusive counterpart $D/L_0$.

To quantify the asymmetry in attachment/detachment as adatoms approach steps from the upper or lower terrace, we introduce
\begin{equation} \label{eq:S}
S:=\frac{\kappa_+}{\kappa_-},
\end{equation}
so that $S\in(0,1)$ corresponds to an inverse Ehrlich--Schwoebel (iES) barrier, $S>1$ to a direct Ehrlich--Schwoebel (ES) barrier, and $S=1$ to the absence of any such barrier. 

Finally, the strength of elastic interactions between steps is measured by the dimensionless coefficients
\begin{equation} \label{eq:oaob}
\oa:=\frac{a^2\alpha}{k_{_B} T L_0^3} \quad \mbox{and} \quad \ob:=\frac{a^2\beta}{k_{_B} T L_0},
\end{equation}
and the dimensionless kinetic modulus is given by
\begin{equation}
\overline{b}:=\frac{L_0 k_{_B} T b}{a^2 D}.
\end{equation}

Let $x$ and $t$ be the nondimensional space and time variables. The dimensionless moving-boundary problem is now given by
\begin{empheq}[left=\empheqlbrace]{align} \label{eq:fbvp2}
\begin{aligned}
 \partial_t \rho_n &= \partial_{xx}\rho_n -\onu \rho_n + \oF ,\\[4pt]
-\rho_n^- \dot{x}_{n+1} - (\partial_x\rho_n)^-&=\underbrace{\ok \left( \rho_n^- - 1  - \chi \Theta \llbracket \rho \rrbracket_{x_{n+1}} + \mathfrak{f}_{n+1} +\frac{\dot{x}_{n+1}}{\overline{b}}\right)}_{J_{n+1}^-} - \okp \llbracket \rho \rrbracket _{x_{n+1}},\\[4pt]
\rho_n^+ \dot{x}_{n} + (\partial_x\rho_n)^+&=\underbrace{\ok S \left( \rho_n^+ - 1 - \chi \Theta \llbracket \rho \rrbracket_{x_{n}} + \mathfrak{f}_{n}+\frac{\dot{x}_n}{\overline{b}}\right)}_{J_{n}^+} + \okp \llbracket \rho \rrbracket_{x_{n}},\\
\dot{x}_n&=\Theta (J_{n}^++J_{n}^-),
\end{aligned}
\end{empheq}
where $\rho_n$, $x_n$, and $x_{n+1}$ now denote the nondimensional adatom density and step positions, and $\mathfrak{f}_n$ is rewritten in dimensionless form as
\begin{equation} \label{eq:fn2}
\mathfrak{f}_n:=\sum_{\substack{r\in\{-R,...,R\} \\ r \neq 0 } } \left\{\frac{\ob}{x_{n+r}-x_n} - \frac{\oa}{(x_{n+r}-x_n)^3}\right\}.
\end{equation}

Recall that the term containing $\overline{b}$ in \eqref{eq:mus} is associated to the dissipation that results from the nonequilibrium processes underlying step migration. In light of the available experimental data, we are unable to estimate $b$, and thus $\overline{b}$. However, since we have restricted our attention to small departures from equilibrium, we shall only consider the limit $\overline{b} \rightarrow \infty$ (which, if the step velocity as given by the kinetic relation is to remain finite, means that the driving force $\digamma$ is vanishingly small), thus making the contribution of $\overline{b}$ to \eqref{eq:fbvp2}$_{2,3,4}$ negligible.

\subsection{Quasistatic approximation}
\label{sub:approx}

Under the quasistatic approximation, the transient term $\partial_t \rho_n$ is neglected in \eqref{eq:fbvp2}$_1$,  as are the advective
terms $\rho_n^{-} \dot{x}_{n+1}$ in \eqref{eq:fbvp2}$_{2}$ and $\rho_n^{+} \dot{x}_{n}$ in \eqref{eq:fbvp2}$_{3}$, thus reducing \eqref{eq:fbvp2} to
\begin{empheq}[left=\empheqlbrace]{align} \label{eq:fbvpQS}
\begin{aligned}
0&= \partial_{xx}\rho_n - \onu \rho_n + \oF ,\\[4pt]
- (\partial_x\rho_n)^-&=\ok ( \rho_n^- - 1 - \chi \Theta(\rho_{n+1}^+-\rho_n^-)
+ \mathfrak{f}_{n}) - \okp(\rho_{n+1}^+-\rho_n^-),\\[4pt]
 (\partial_x\rho_n)^+&=\ok S ( \rho_n^+ - 1 - \chi \Theta (\rho_{n}^+-\rho_{n-1}^-)
+ \mathfrak{f}_{n}) + \okp (\rho_{n}^+-\rho_{n-1}^-),\\[4pt]
\dot{x}_n&=\Theta (J_{n}^++J_{n}^-).
\end{aligned}
\end{empheq}

In the literature on step flow \citep{Krug2005,Michely2012}, this approximation is presented as appropriate in the regimes of slow deposition or evaporation
\begin{equation}\label{eq:qsacond}
\oF \Theta \ll 1 \quad \mbox{or} \quad \onu \Theta \ll 1.
\end{equation}
As explained in \citet{Michely2012}, \eqref{eq:qsacond} is always satisfied in the step-flow regime since, were it violated, crystal growth would occur in a different regime, one that involves island nucleation. However, that \eqref{eq:qsacond} indeed permits to carry out the stability analysis on the quasistatic system \eqref{eq:fbvpQS} without missing important effects on the stability of steps is not substantiated. Hence, we rather consider the quasistatic approximation as an \emph{a priori} simplification, one that requires \emph{a posteriori} checking, as we shall do in the sequel paper via a linear-stability analysis of the moving-boundary problem \eqref{eq:fbvp2}. 

\section{Stability analysis}
\label{sec:stability1}

We start in Section~\ref{sub:sss} by computing the steady-state solution of \eqref{eq:fbvpQS} corresponding to a train of equidistant steps that migrate at constant velocity and identical adatom distributions on all terraces. The method underlying the linear-stability analysis of this fundamental solution is presented next in Section~\ref{sub:method}, in which the moving-boundary problem is reduced to a dynamical system for the vector-valued perturbation and then solved by means of the Fourier transform to yield the dispersion relation. Finally Section~\ref{sub:reduction} is devoted to a conjecture that infers the stability of all perturbations from that of step-pairing and long-wavelength perturbations.

\subsection{Steady-state solution} 
\label{sub:sss}

Consider an infinite sequence of steps assumed all equidistant at $t=0$, and let the initial position of the $n$th step be $x_n(0)=n$ ($n \in \mathbb{Z}$). The principal solution of \eqref{eq:fbvpQS} consists of an adatom density $\rz(x,t)$, defined everywhere except at steps where it is dicontinuous, and steps that propagate at the same speed $\vz$, so that the position of the $n$th step at time $t$ is given by $\xz_n(t)=n+\vz t$. Since the adatom density is the same on all terraces, $\rz(x,t)$ is given by
\begin{equation}
\rz(x,t)=\rtz(x-\xz_n(t)) \quad \forall  x \in(\xz_n(t),\xz_n(t)+1),
\end{equation}
with $\rtz$ defined on $(0,1)$.

Denoting by $\rz^+$ and $\rz^-$ the limiting values of $\rtz$ at $0^+$ and $1^-$, integration of \eqref{eq:fbvpQS}$_1$ yields the adatom profile on each terrace up to two arbitrary constants $\rz^+$ and $\rz^-$. These are computed through the boundary conditions \eqref{eq:fbvpQS}$_{2,3}$, hence completely determining the adatom distribution on all terraces. Finally the velocity $\vz$ of the steps is computed via \eqref{eq:fbvpQS}$_{4}$.\footnote{As a result of the preservation of equidistance between steps in this fundamental solution, $\mathfrak{f}_n=0$.} 

The analytical expressions for $\rtz(x)$ and $\vz$ are quite lengthy in the general case involving both deposition and sublimation; for the sake of clarity of the presentation, we do not report them. Instead, we display the expression of  $\rtz(x)$ when evaporation is negligible ($\onu=0$):
\begin{equation} \label{eq:rhoeq}
\rtz(x)=-\frac{1}{2}\oF x (x-1) + (\rz^- - \rz^+) x + \rz^+,
\end{equation}
where, letting $A:=S-1$ and $B:=\ok S +\okp S +\okp+S+1>0$,
\begin{equation}
\rz^+=1+\frac{\oF(\ok (1- \Theta A)+2 \okp +2)}{2 \ok B} \quad \mbox{and} \quad \rz^-=1+\frac{\oF \big( \ok (S- \Theta A)+2 \okp +2)}{2 \ok B}.
\end{equation}
The resulting step velocity is given by
\begin{equation}
\vz=\oF \Theta.
\end{equation}
Note that, in the absence of the Ehrlich--Schwoebel effect ($S=1$), $\rz^+=\rz^-$, rendering $\rtz$ symmetric with respect to $x=1/2$. Equation \eqref{eq:rhoeq} allows us to estimate the maximum departure of adatom density to its equilibrium value under deposition and thereby specify the conditions of validity of \eqref{eq:linear}. Assuming\footnote{%
This is a reasonable assumption as can be seen from the quantitative estimates of $S$ provided in Appendix~C of Part~II.}
 that $0.1<S<10$, $\max_{x\in(0,1)} \big\vert \rtz(x)-1\big\vert \sim \max(\oF/8,\oF/\ok)$. Hence, the assumption \eqref{eq:linear} is satisfied as long as $\oF \ll 10$ and $\oF \ll \ok$. The first condition is mostly fulfilled by virtue of \eqref{eq:qsacond} (since typically $\Theta \sim 0.01$ to $0.1$) while the second, more restrictive when $\ok <1$, needs to be checked. Similarly, one can show that under evaporation $\max_{x\in(0,1)} \big\vert \rtz(x)-1\big\vert \sim \max(\onu/8,\onu/\ok)$, furnishing, for \eqref{eq:linear} to be satisfied, the analogous conditions $\onu \ll 10$ and $\onu \ll \ok$. 

\subsection{Linear stability analysis 
\label{sub:method}}

To investigate the stability of the fundamental solution against bunching, we introduce the displacement $\zeta_n(t):=x_n(t)-\xz_n(t)$ of the $n$th step away from its position in the train of equidistant steps. We write the adatom density on $n$th terrace in the form
\begin{equation}
\rho_n(x,t)=\rt_n(x- \xz_n(t)) \quad \forall \, x \in \big( x_n(t), x_{n+1}(t)\big),
\end{equation}
with $\rt_n$ defined on the interval $(\zeta_n(t),1+\zeta_{n+1}(t))$. Integration of  \eqref{eq:fbvpQS}$_1$ yields an expression of $\rt_n$ in terms of the displacements of the steps that bound the $n$th terrace, $\zeta_n(t)$ and $\zeta_{n+1}(t)$, and the associated limiting values, $\rho_n^+(t):=\rt_n\big(\zeta_n(t),t \big)$ and   $\rho_n^-(t):=\rt_n\big(1+\zeta_{n+1}(t),t\big)$, which we formally write as 
\begin{equation}
\rho_n(x,t)=\check{\rho}[\rho_n^-(t),\rho_n^+(t),\zeta_n(t),\zeta_{n+1}(t)](x),
\end{equation}
where $\check{\rho}$ is a known function whose analytical expression we do not write explicitly in the general case. Instead, when growth takes place without evaporation ($\onu=0$), $\check{\rho}$ reads
\begin{multline}
\check{\rho}[\rho_n^-(t),\rho_n^+(t),\zeta_n(t),\zeta_{n+1}(t)](x) =-\frac{\oF}{2}(x-\zeta_n(t))(x-\zeta_{n+1}(t) -1)\\+ \frac{\rho_n^-(t)(x- \zeta_n(t)) -\rho_n^+(t)(x-\zeta_{n+1}(t)-1)}{\zeta_{n+1}(t)+1-\zeta_n(t)}.
\end{multline}

The remaining conditions \eqref{eq:fbvpQS}$_{2,3,4}$ yield a first order dynamical system of three equations for the three scalar unknowns associated to the $n$th step: $\rho_{n}^-(t)$, $\rho_n^+(t)$ and $\zeta_n(t)$. Letting $\mathbf{p}_n(t):=\{\rho_{n}^-(t), \rho_{n}^+(t), \zeta_n(t)\}$, this system can be formally written as
\begin{equation} \label{eq:sd}
\mathbf{M} \dot{\mathbf{p}}_{n}
= \bm{\mathcal{F}}\big(\mathbf{p}_{n-R},..., \mathbf{p}_{n+R+1} \big),
\end{equation}
where the superposed dot denotes differentiation with respect to time, 
\begin{equation}
\mathbf{M}:= \begin{pmatrix}
0 & 0& 0 \\
0 &0 &0 \\
0 & 0& 1 
\end{pmatrix},
\end{equation}
and $\bm{\mathcal{F}}$ is the vector function with components
\begin{empheq}[left=\empheqlbrace]{align} \label{eq:functionf}
\begin{aligned}
\cF_1(\mathbf{p}_{n-R},..., \mathbf{p}_{n+R+1}):=
 & \frac{\mathrm{d} \check{\rho}}{\mathrm{d}x} \bigg\rvert_{x=\zeta_n}+J_{n+1}^-
 -\okp (\rho_{n+1}^+-\rho_{n}^- ), \\[4pt]
\cF_2(\mathbf{p}_{n-R},..., \mathbf{p}_{n+R+1}):= 
&\frac{\mathrm{d} \check{\rho}}{\mathrm{d}x} \bigg\rvert_{x=1+\zeta_{n+1}}-J_n^+-\okp (\rho_{n}^+-\rho_{n-1}^- ), \\[4pt]
\cF_3(\mathbf{p}_{n-R},..., \mathbf{p}_{n+R+1}):=
 & \Theta( J_{n}^+ + J_{n}^-)-\vz.
\end{aligned}
\end{empheq}
In \eqref{eq:functionf}, 
\begin{equation}
\begin{aligned}
J^-_{n}&:= \ok(\rho_{n-1}^- - 1- \chi \Theta (\rho_{n}^+-\rho_{n-1}^- ) + \mathfrak{f}_{n}),\\[4pt]
J^+_{n}&:=\ok S (\rho_n^+ - 1- \chi \Theta (\rho_{n}^+-\rho_{n-1}^- ) + \mathfrak{f}_{n}),
\end{aligned}
\end{equation}
with $\mathfrak{f}_n$ rewritten in terms of the step displacements:
\begin{equation} \label{eq:fnzeta}
\mathfrak{f}_n=\sum_{\substack{r\in\{-R,...,R\} \\ r \neq 0}} \left\{\frac{\ob}{\zeta_{n+r}-\zeta_n+r} - \frac{\oa}{(\zeta_{n+r}-\zeta_n+r)^3}\right\}.
\end{equation}

Next, we linearize the system \eqref{eq:sd}--\eqref{eq:fnzeta} about the principal solution $\pz_n := \{\rz^-,\rz^+,0\}$. Denoting by $\delta \mathbf{p}_n(t) :=\{\delta \rho_{n}^-(t),\delta \rho_{n}^+(t),\delta \zeta_n(t)\}$ the perturbation, we thus obtain the linear system
\begin{equation} \label{eq:linper}
\mathbf{M} \delta \dot{\mathbf{p}}_{n}= \sum_{r=-R}^{R+1} \frac{\partial \bm{\mathcal{F}}}{\partial \mathbf{p}_{n+r}} \bigg \rvert_{\pz} \delta \mathbf{p}_{n+r}.
\end{equation}

This system can be diagonalized using the spatial Fourier transform of $\{\delta \mathbf{p}_n\}_{n \in \mathbb{Z}}$
\begin{equation}
\widehat{\delta \mathbf{p}}_k(t):=\sum_{n=-\infty}^{+\infty} \delta \mathbf{p}_n(t) \me^{i k n},
\end{equation}
where $k$ is the wavenumber indexing the Fourier modes and $2 \pi /k$ the associated wavelength\footnote{All possible wavelengths are accounted for by taking  $k$ in the first Brillouin zone $[-\pi , \pi]$, and an arbitrary perturbation $\delta \mathbf{p}_n(t)$ can be expressed as
$ \delta \mathbf{p}_n(t) =(2\pi)^{-1} \int_{-\pi}^{\pi} \widehat{\delta \mathbf{p}}_k(t) \me^{-i k n} \ud k.$}.
The Fourier transform of \eqref{eq:linper} reads
\begin{equation} \label{eq:linperhat}
\mathbf{M}\dot{ \widehat{\delta \mathbf{p}}}_{k} = {\underbrace {\Bigg( \sum_{r=-R}^{R+1} \frac{\partial \bm{\mathcal{F}}}{\partial \mathbf{p}_{n+r}} \bigg \rvert_{\pz} \me^{i k r} \Bigg) }_{\mathbf{C}_k}} \widehat{\delta \mathbf{p}}_{k},
\end{equation}
where $\mathbf{C}_k$ denotes the $3\times3$ matrix associated to the $k$th Fourier mode.\footnote{For a slightly different perspective, consider the linear-perturbation problem for a finite number of steps, $N$, with periodic boundary conditions. The right-hand side of \eqref{eq:linper}, written for $n=0, \ldots ,N-1$, has the form of a linear system for the $3N$ variables $\{\delta \rho_{0}^-(t),  \delta \rho_{0}^+(t),\delta \zeta_0(t),\ldots ,\delta \rho_{N-1}^-(t), \delta \rho_{N-1}^+(t),\delta \zeta_{N-1}(t)\}$. The associated $3N \times 3N$ matrix has a block-circulant architecture and $\mathbf{C}_k$, as given in \eqref{eq:linperhat}, is the matrix-eigenvalue that derives from its block diagonalization.}

Under the condition that the determinant of the upper-left $2 \times 2$ submatrix of $\mathbf{C}_k$ is not zero, a condition that can be shown to hold as long as the adatom equilibrium coverage satisfies $\Theta <0.5$, the linear system consisting of the first two rows of \eqref{eq:linperhat} can be solved to express $\widehat{\delta \rho}^-_k$ and $\widehat{\delta \rho}^+_k$ in terms of $\widehat{\delta \zeta}_k$, so that the third row of \eqref{eq:linperhat} can be rewritten as
\begin{equation}
\dot{\widehat{\delta \zeta}}_k=\lambda(k) \widehat{\delta \zeta}_k,
\end{equation}
with $\lambda(k)$ the sought-after dispersion relation. 

Thus, the time evolution of an initial perturbation containing the sole mode of wavenumber $k$, with shape $\delta \zeta_n(0)=\widehat{\delta \zeta}_k(0) \me^{-ikn}$, is given by
\begin{equation}
\delta \zeta_n(t)=\widehat{\delta \zeta}_k(0) \exp[i\, (\mathrm{Im}(\lambda(k)) t-kn)+\mathrm{Re}(\lambda(k)) t].
\end{equation}
Clearly, the train of equidistant steps is linearly stable with respect to bunching as long as  $\mathrm{Re}(\lambda(k))<0$ for all $k \in [-\pi,\pi]$. Since $\lambda(-k)=\overline{\lambda(k)}$, $\mathrm{Re}(\lambda(k))$ needs only to be studied on $(0, \pi]$.

In summary, the procedure detailed above yields an analytical expression for the dispersion relation $\lambda(k)$, with the stability of the steady-state solution, or lack thereof, determined by the sign of its real part.  We do not report the lengthy expression of $\mathrm{Re}(\lambda(k))$ in the general case.

\subsection{Reduction to long-wavelength and step-pairing perturbations}
\label{sub:reduction}

\begin{figure*}[tb]
\begin{center}
\includegraphics[width=87mm]{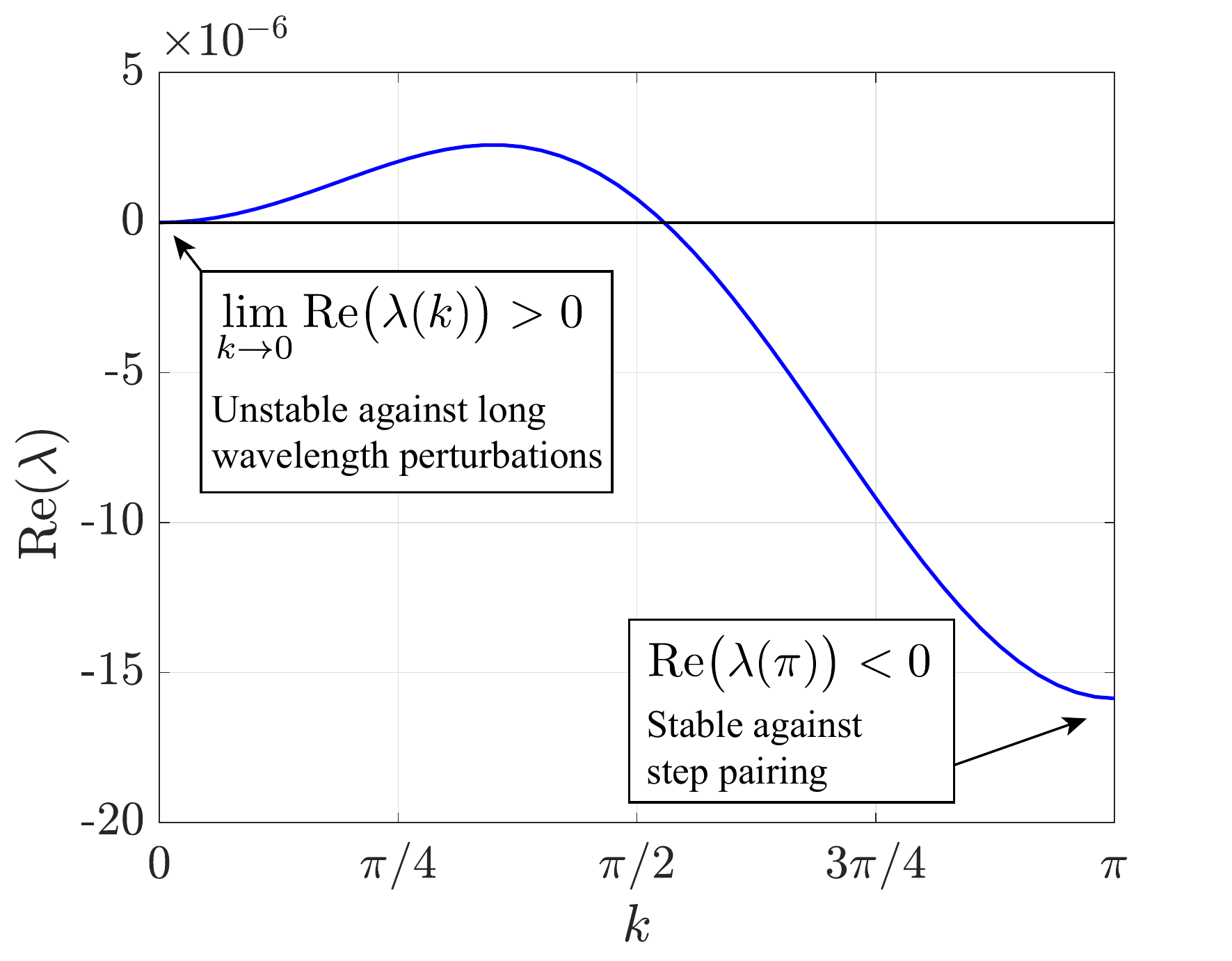}
\caption{\label{fig:dc1} An example of the dispersion curve in the quasistatic deposition regime ($\onu=0$), with $\oF=10^{-1}$, $S=0.4$, $\ok=20$, $\Theta=0.01$, $\oa=10^{-4}$, and $\ob=0$. Steps are unstable against long-wavelength perturbations ($k \rightarrow 0$) and stable with respect to step pairing ($k=\pi$).}
\end{center}
\end{figure*}

A typical dispersion curve is on display in Fig.~\ref{fig:dc1}. While the signs of $\displaystyle \lim_{k \rightarrow 0} \mathrm{Re}(\lambda(k))$ and $\mathrm{Re}(\lambda(\pi))$ control the stability against long-wavelength perturbations and step pairing, respectively, $\mathrm{Re}(\lambda(k))$ for an arbitrary value of $k$ in $(0, \pi]$ determines the stability, or lack thereof, against perturbations of intermediate wavelength ${2\pi}/{k}$. However, an extensive numerical study of the dispersion curves for physically relevant values of the model's dimensionless parameters, as estimated in Appendix~C of the sequel paper from the experimental literature on the homoepitaxy of GaAs(001) and Si(111)-$7\!\times\!7$ as well as the heteroepitaxy of SiGe on Si(001), leads us to the following

\begin{conj} \label{prop:reduction}
If $\displaystyle Re(\lambda(k)) <0$ for $k \ll 1$ and $Re(\lambda(\pi))<0$, then $Re(\lambda(k))<0$ for all $k\in(0,\pi]$.
\end{conj}
In other words, it is enough to check the stability of steps against both long-wavelength and step-pairing perturbations to ensure their stability for all wavelengths. Conversely, it is clear that the instability of either one of these two limit perturbations is sufficient for step flow to be unstable.\footnote{We note that Conjecture \ref{prop:reduction} is not valid in the very particular case where both attractive and repulsive elastic interactions are present, of comparable magnitudes, and with range $R$ larger than 1 (i.e., when interactions beyond nearest neighbors are considered). In such a situation, instabilities of finite wavelength may develop even though long-wavelength and step-pairing perturbations are both stable.}

\paragraph{Proof of Conjecture \ref{prop:reduction} for the case $\chi=0$}

We formally take $\chi=0$ and restrict our attention to nearest-neighbor interactions ($R=1$). The validity of the latter assumption is discussed in Section \ref{sub2:range} below; the former simplification has no physical basis, as discussed in Section~\ref{sub:thermo} above.\footnote{However, by taking $\chi=0$ in \eqref{eq:fbvp2} one recovers the equations that govern step flow as written in \citet{Pierre-Louis2003}.} Both are introduced in order to obtain an analytically tractable dispersion relation. Recalling that $A:=S-1$ and $B:=\ok S + \okp S + S +\okp + 1$, this relation now reads
\begin{equation} \label{eq:depoNoTC}
\mathrm{Re}( \lambda(k))= \sin^2(k/2) \left\{ \frac{2 \ok \Theta \big( -A \oF (1+S) +8 C (\ob-3\oa ) \sin^2 (k/2) \big)}{B \big(4 \okp \sin^2(k/2)+\ok B\big)} \right\},
\end{equation}
where $C:=\okp (S+1)(2\ok S+S+1)+\ok S (\ok S+S+1) + \okp^2(S+1)^2$. Since the denominator of  \eqref{eq:depoNoTC} is strictly positive, the sign of  $\mathrm{Re}\big( \lambda(k)\big)$ coincides with that of its numerator. The latter has the form $a \sin^2(k/2)+b \sin^4(k/2)$, with $a:=-2 \ok \Theta A \oF (1+S)$ and $b:=16 \ok \Theta C (\ob-3\oa )$. The assumption that both long-wavelength and step-pairing instabilities are stable implies that $a<0$ and $a+b<0$. These conditions are sufficient for $\mathrm{Re}(\lambda(k))$ to be negative for all $k \in (0,\pi]$.

\section{Results}   
\label{sub2:mech}

Hereafter, we deploy a mostly analytical approach to understand how the physics of step flow determines the onset of the bunching instability. In Section~\ref{subsec:mrp}, we introduce a classification in terms of mechanisms, operating regimes, and material parameters. 
This is followed by a study of the stabilizing or destabilizing character of each of the mechanisms under consideration in the deposition 
(Section~\ref{subsec:deposition}) and sublimation (Section~\ref{subsec:sublimation}) regimes. Since stability is determined by the competition between mechanisms that are either stabilizing or destabilizing, Section~\ref{sub2:interplay} investigates how the various operating and material parameters govern the interplay between mechanisms. Finally, we examine in Section~\ref{sub2:range} the validity of the approximation of nearest-neighbor interactions by looking at how the dispersion curve changes with elastic interactions of longer ranges.

\subsection{Mechanisms, regimes, and parameters}
\label{subsec:mrp}

\begin{figure*}[tb]
\begin{center}
\includegraphics[width=0.4\textwidth]{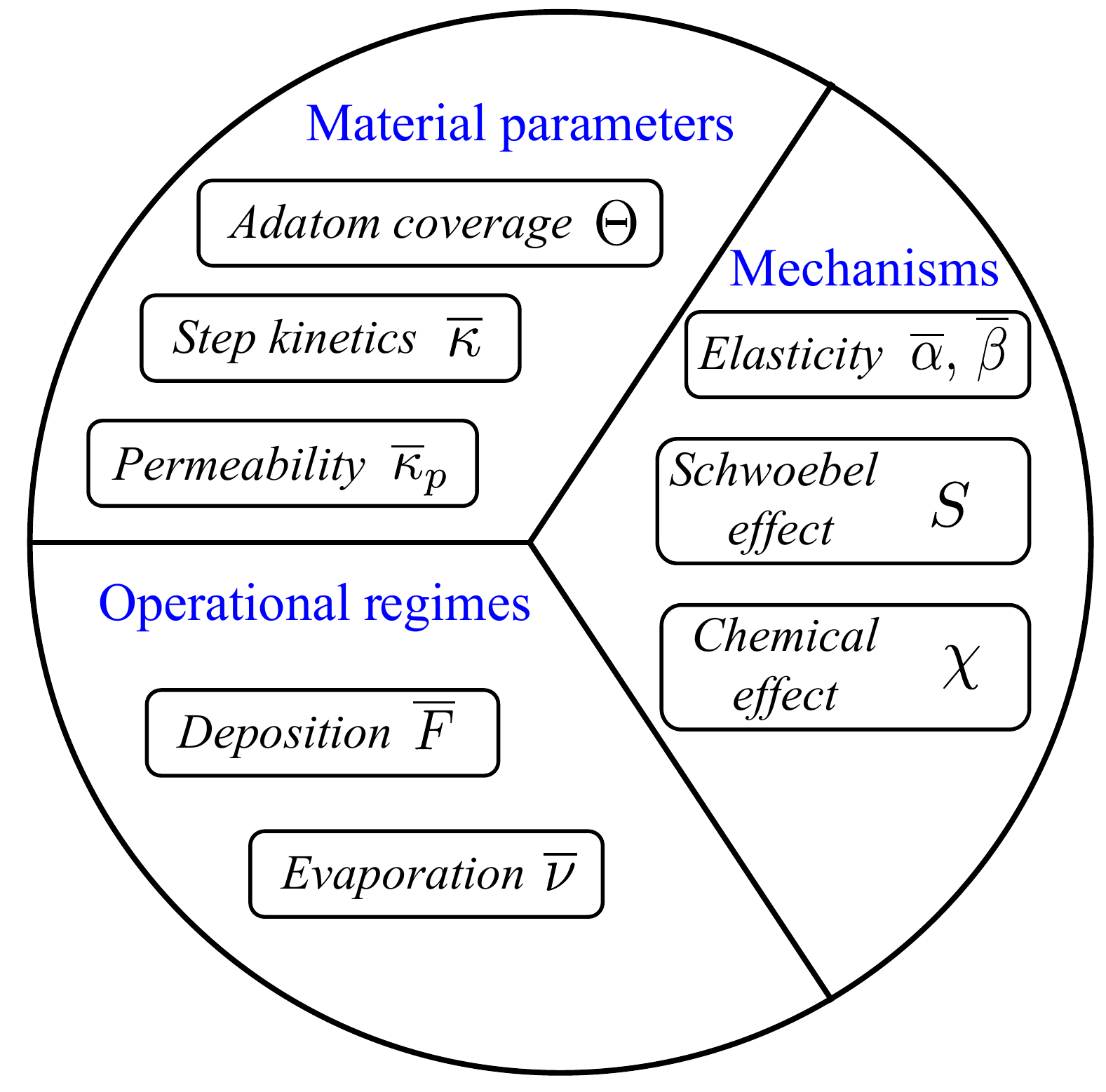}
\caption{Diagram of the physical factors that govern the stability of steps against bunching. We distinguish the mechanisms that are intrinsically stabilizing or destabilizing from the parameters, material or operating, that determine the relative importance of the aforementioned mechanisms.}
\label{fig:mech} 
\end{center}
\end{figure*}

As sketched in Fig.~\ref{fig:mech}, we distinguish between operating regimes, mechanisms which are stabilizing or destabilizing, and material parameters whose variations alter the relative influence of these mechanisms on the stability of steps against bunching:
\begin{itemize}

\item[--] \textit{Mechanisms}. These are the physical processes that tend to stabilize or destabilize steps, namely, the (dipole-dipole and monopole-monopole) \textit{elastic interactions} between steps, the \textit{Ehrlich--Schwoebel barrier}, and the \textit{chemical effect} which couples the diffusion fields on adjacent terraces and accounts for the contribution to driving force at steps of neighboring adatoms.\footnote{In Part~II of this paper, in which the linear-stability analysis is extended beyond the quasistatic approximation, the \textit{dynamics effect} is added to the list of mechanisms.}

\item[--] \textit{Operating regimes}. By which we mean \emph{deposition} and \emph{evaporation}, considered independently. Two classes of mechanisms can be distinguished. On one hand, elasticity whose influence on stability is independent of the deposition or evaporation rate; this is an \emph{energetic} mechanism. On the other hand, the remaining mechanisms, whose influence on stability grows linearly with the deposition or evaporation rate; we refer to these mechanisms as \emph{kinetic}.

\item[--] \textit{Material parameters}. These parameters, which determine the relative influence on stability of the different mechanisms involved, are the \emph{adatom equilibrium coverage} $\Theta$, the dimensionless \emph{adatom attachment/detachment coefficient} $\ok$, and the \emph{step permeability} $\okp$. Since stabilizing/destabilizing mechanisms are simultaneously present, the dominant mechanism depends on these material parameters.

\end{itemize}

\subsection{Deposition}
\label{subsec:deposition}

When adatom desorption is negligible ($\onu=0$) and elastic interactions between steps are restricted to nearest neighbors ($R=1$), the growth rate $\mathrm{Re}(\lambda_{SP}^{dep})$ of the step-pairing instability ($k=\pi$) is given by
\begin{equation} \label{eq:depoSP}
\mathrm{Re}(\lambda_{SP}^{dep})= \frac{\Theta \ok(2 \oF (S+1)(2 B \chi \Theta-A)+16 C (\ob-3\oa))}{B( \ok (B-2 \chi \Theta A) + 4 \okp)},
\end{equation}
For long-wavelength perturbations in the same regime, a series expansion of $\mathrm{Re}(\lambda(k))$ about $k = 0$ yields
\begin{equation} \label{eq:depoLW1}
\mathrm{Re}(\lambda(k))=-\frac{\oF \Theta (S+1) A}{2 B^2}k^2+o(k^2)
\end{equation}
in the presence of the Ehrlich--Schwoebel barrier or its inverse ($S\neq1$), and 
\begin{equation} \label{eq:depoLW2}
\mathrm{Re}(\lambda(k))=\frac{\Theta (\ok+2 \okp)( (\ok+2) \oF \chi \Theta -2 \ok (\ok + 2 \okp +2) (3 \oa - \ob))}{2 \ok (\ok + 2 \okp +2 )^2} k^4+o(k^4)
\end{equation}
when adatom attachment/detachment at steps is symmetric ($S=1$). In \eqref{eq:depoSP}--\eqref{eq:depoLW2}, $A:=S-1$, $B:=\ok S + \okp S + S +\okp + 1$, and $C:=\okp (S+1)(2\ok S+S+1)+\ok S (\ok S+S+1) + \okp^2(S+1)^2$.

Consider first step-pairing. When $ \Theta <0.5$, a condition which is always satisfied in practice, the denominator of \eqref{eq:depoSP} remains positive. By examining the sign of each term in the numerator, we see that the Ehrlich--Schwoebel barrier, which manifests itself through $A:=S-1$, is stabilizing when it is direct and destabilizing when it is inverse; the elastic monopole-monopole interactions, manifest through $\ob$, are destabilizing, whereas their dipole-dipole counterparts, manifest through $\oa$, are stabilizing; the adatom jump effect, manifest through $\chi$, is destabilizing.

Next, consider long-wavelength perturbations. As can be seen from \eqref{eq:depoLW1}, it is the Ehrlich--Schwoebel barrier which, when present, determines the stability of steps against these perturbations; as with step-pairing, the direct ES barrier stabilizes against bunching whereas its inverse is destabilizing. When adatom attachment/detachment is symmetric, \eqref{eq:depoLW2} shows that the stability is governed by the interplay between the elastic step-step interactions and the chemical effect, with the stabilizing or destabilizing influence of each of these two mechanism the same as for step-pairing. 

Importantly, we see from \eqref{eq:depoSP}--\eqref{eq:depoLW2} that step permeability is not a stabilizing or destabilizing mechanism \emph{per se}, in the sense that if other mechanisms are disabled (by setting $S=1$, $\chi =0$, and $\oa=\ob=0$) the permeability of steps has no effect on their stability against either pairing or long-wavelength perturbations. Instead, what permeability does is to change the relative weights of the stabilizing or  destabilizing mechanisms under consideration. For example, in the limit of transparent steps ($\okp \rightarrow \infty$), it can be seen from the numerators of \eqref{eq:depoSP} and \eqref{eq:depoLW2} that elastic interactions become predominant over the Ehrlich--Schwoebel barrier and the chemical effect.\footnote{Indeed, in both in \eqref{eq:depoSP} and \eqref{eq:depoLW2}, the elastic terms are multiplied by prefactors with larger powers of $\okp$ than those of the remaining terms.} This is expected on physical grounds, since having very permeable steps amounts to shutting the asymmetry in the adatom attachment/detachment and equalizing their densities on both sides of each step, thus minimizing the influences of either the direct ES barrier or its inverse and of the chemical effect.

Finally, an interesting conclusion that can be drawn from \eqref{eq:depoSP}--\eqref{eq:depoLW2} is that the different mechanisms do not exhibit a complex interplay, in the sense that for any set of parameters the  influence on stability of each mechanism remains the same. Indeed, the effects of the various mechanisms are essentially additive, and changing the parameters modifies their relative weights but not the signs of their prefactors.

\subsection{Evaporation}
\label{subsec:sublimation}

Consider now the evaporation regime ($\oF=0$). The growth rate of the step-pairing instability is given by
\begin{equation} \label{eq:lambdaEvap}
\mathrm{Re}(\lambda_{SP}^{dep})=\frac{8 \sn \exp(2\sn) \ok \Theta( N_1+(\ob-3\oa) N_2)}{\mathfrak{D}},
\end{equation}
where, letting $E:=\okp+(\ok +\okp)S$, the denominator has the form
\begin{equation}
\begin{aligned}
\mathfrak{D}:= &\, 4 \exp(2 \sn)\bigg\{ \sn\Big[-2 \okp + \ok (S-1) \chi \Theta +(2 \okp +\ok (S (1-\chi \Theta)+1+\chi \Theta))\cosh\big(\sn\big) \Big]\\
&\, +(\onu + \ok D) \sinh\big(\sn\big)\bigg\}\bigg\{\sn \Big[ 2 \okp + \ok (1-S) \chi \Theta + (2 \okp +\ok (S (1-\chi \Theta)+1+\chi \Theta))\cosh\big(\sn\big)\Big]\\
&\, +(\onu + \ok E) \sinh(\sn)\bigg\}.
\end{aligned}
\end{equation}
In \eqref{eq:lambdaEvap}, the kinetic term contribution $N_1$ and its elastic counterpart  $N_2$ of the numerator read
\begin{empheq}[left=\empheqlbrace]{align} 
\begin{aligned}
\label{eq:N1}
N_1:= & (1+S)\ok\big\{\onu^{3/2}A-\chi \Theta {\onu}\big[\sn(1+S)(1+\cosh(\sn))+2 E \sinh(\sn)\big]\big\}, \\[4pt]
N_2:= & 8 \sn \exp(2 \sn)  \ok \Theta\big[ N_{21}+N_{22} \cosh(\sn)+N_{23} \cosh (2 \sn)+N_{24} \sinh(\sn)
+N_{25} \sinh(2\sn)\big], \\
\end{aligned}
\end{empheq}
where
\begin{empheq}[left=\empheqlbrace]{align} \label{eq:functionf2}
\begin{aligned}
N_{21}&:=-\sn E[4 \okp-\ok (2 \chi \Theta  A +S+1)]-\onu\sn(S+1), \\[4pt]
N_{22}&:= 4 \ok\sn(1+S) E,\\[4pt]
N_{23}&:= \sn\big\{E[4 \okp+\ok (3+3 S+2 (1-S) \chi \Theta)]+(1+S) \onu\big\}, \\[4pt]
N_{24}&:= 2 \ok[2 E^2+2 S\onu +(S^2-1) \chi \Theta \onu], \\[4pt]
N_{25}&:= 4 \onu \okp (1+S)+2 \ok E^2+\ok[1+\chi \Theta+S (4+S-S \chi \Theta)] \onu.
\end{aligned}
\end{empheq}
Assuming that $\Theta <0.5$ (an experimentally sound assumption), one can show that $N_2>0$ and since $\cosh(\sn)>1$, one can also show that $\mathfrak{D}>0$.

Since $N_2>0$, the elastic interactions have the same effect on stability as under deposition, i.e., monopole-monopole interactions are destablizing whereas dipole-dipole interactions are stabilizing. This is not the case for the remaining mechanisms. Recalling that $A:=S-1$, it follows from \eqref{eq:N1} that the direct Ehrlich--Schwoebel barrier is now destabilizing, whereas the inverse ES barrier has become stabilizing. Similarly, by noticing that the term that multiplies $\chi$ in \eqref{eq:N1} is positive, we conclude that the chemical effect is now stabilizing against bunching, in contrast to its effect under deposition.

Moreover, as with deposition, we note the absence of complex interplay between the various mechanisms under consideration. Indeed, changing the material parameters alters the relative importance of these mechanisms but not the signs of their prefactors, so that the stabilizing or destablizing character of each is unchanged. Likewise, step permeability is neither a stabilizing nor a destabilizing mechanism in itself. 
\begin{table*}[tb]
\begin{center}
\begin{tabular}{l c c c c c}
\hline
& ES & iES & CE & DDE & MME \\
& $S>1$ & $S<1$ & $\chi$ & $\oa$ & $\ob$ \\
\hline
\hline
Deposition & $\mathcal{S}$ & $\mathcal{D}$ & $\mathcal{D}$ & $\mathcal{S}$ & $\mathcal{D}$\\
\hline
Evaporation &  $\mathcal{D}$ & $\mathcal{S}$ &$\mathcal{S}$  & $\mathcal{S}$ & $\mathcal{D}$ \\
\hline
\end{tabular}
\caption{Effects of each of the basic mechanisms on the onset of the bunching instability.  ES refers to the Ehrlich--Schwoebel barrier, iES to its inverse, CE to the chemical effect, DDE to dipole-dipole elastic interactions, and MME to their monopole-monopole counterparts.  $\mathcal{S}$ stands for stabilizing and $\mathcal{D}$ for destabilizing.}
\label{tab:mech}
\end{center}
\end{table*}

Because of the complexity of the dispersion relation for long-wavelength perturbations, an analytical study is not feasible. Instead, numerics confirm that each mechanism has qualitatively the same effect on stability as for step-pairing.
The effects of each of the three mechanisms are summarized in Table~\ref{tab:mech}.

\subsection{Scalings with the operating and material parameters }
\label{sub2:interplay}

The onset of step bunching in the presence of several intrinsically stabilizing or destabilizing mechanisms is governed by their interplay. Based on the expressions \eqref{eq:depoSP}--\eqref{eq:lambdaEvap} for the growth rate of step-pairing and long-wavelength perturbations, we now discuss the influence of the parameters of the model on the relative weights of the different mechanisms. For clarity, we distinguish the parameters that quantify each the strength of a mechanism ($S$ for the ES barrier or its inverse, $\oa$ for the elastic dipole-dipole interactions, and $\ob$ for their monopole-monopole counterparts) from the parameters that govern the interplay between the mechanisms. The latter are either operating ($\oF$ and $\onu$) or material ($\Theta$, $\ok$, and $\okp$).

We begin by noting that the dependence of the growth rate on the deposition and evaporation rates, $\oF$ and $\onu$, leads to a distinction between \textit{kinetic} and \textit{energetic} mechanisms. As seen in the expressions \eqref{eq:depoSP}--\eqref{eq:depoLW2}, the contributions to the growth rate associated with the ES barrier and the chemical effect in the deposition regime are both linear in $\oF$, whereas the contribution associated with the elastic step-step interactions is independent of $\oF$. Under evaporation, a Taylor expansion of \eqref{eq:lambdaEvap} provides an identical dependence on $\onu$ of the contributions to the growth rate associated with the three elementary mechanisms at play. We therefore refer to the Schwoebel barrier and chemical effect as \emph{kinetic} mechanisms and to the elastic step-step interactions as an \emph{energetic} mechanism. As the deposition or evaporation rate increases, the influence of the kinetic mechanisms on the stability of steps against bunching becomes more prominent, at the expense of their energetic counterpart which remains unchanged.

Further, the equilibrium adatom coverage provides a distinction between the chemical effect, whose contribution to $\re(\lambda)$ is quadratic in $\Theta$, and the remaining two mechanisms whose contributions to the growth rate of the perturbation scale linearly with $\Theta$. This implies that the chemical effect is more pronounced for material surfaces with high equilibrium adatom coverage, such as GaAs(001) and Si(111)-1\texttimes1 for which values of $\Theta$ as high as 0.2 have been measured (cf.~Appendix~C of Part~II). 
\begin{table*}[tb]
\begin{center}
\begin{tabular}{c c c c c c}
\hline
&   ES/iES & CE &  E  \\
& $S$ & $\chi$ & $\oa, \, \ob$ \\ 
 \hline
 \hline
 $a$ (for $\oF, \, \onu$) & 1 & 1 & 0  \\
 \hline
 $b$  (for $\Theta$) & 1 & 2 & 1 \\
 \hline
\end{tabular}
\end{center}
\caption{The scalings of the contributions of the basic mechanisms to the growth rates of step-pairing and long-wavelength perturbations with the operating parameters $\oF$ and $\onu$ and the material parameter $\Theta$. The growth rate is additively decomposed into $\textstyle \re(\lambda)=\sum_{m} \re(\lambda_{m})$, with $m$ spanning the three mechanisms: the Ehrlich--Schwoebel barrier or its inverse (ES/iES), the chemical effect (CE), and the elastic step-step interactions (E). The contribution of each mechanism scales as $\re(\lambda_{m}) \propto \oF^a \Theta^b$ under deposition and $\re(\lambda_{mech}) \propto \onu^a \Theta^b$ under sublimation, with the exponents $a$ and $b$ given in the table.}
\label{tab:coeff}
\end{table*}

The scalings of the contributions of the three elementary mechanisms to the growth rate of step-pairing and long-wavelength perturbations with the operating parameters, $\oF$ and $\onu$, and the material parameter $\Theta$ are summarized in Table~\ref{tab:coeff}.

Turning to $\ok$ and $\okp$, we note that since they both account for kinetic processes at the steps, they play comparable roles.\footnote{As noted by \citet{Pierre-Louis2003} and \citet{Sato2000}, permeability can be viewed as a mechanism in parallel with the attachment/detachment of adatoms at steps, by analogy with electric circuits.} Recalling that $\ok$ and $\okp$ are dimensionless velocities associated with the kinetics of adatom attachment/detachment at, and hopping across, steps, $\okp$ has a sensible effect on stability only when $\okp \gtrsim \ok$. This is clear from the expressions \eqref{eq:depoSP}--\eqref{eq:lambdaEvap}, where $\okp$ appears in weighted sums with $\ok$.  Although the scaling of the growth rate with $\ok$ and $\okp$ is not trivial in general, it can be seen in particular cases (such as when the magnitudes of $\ok$, $\okp$, and 1 are clearly separated) that the largest power of $\ok$ and $\okp$ is in the contribution of elasticity to $\re(\lambda)$, with a smaller exponent in the contribution of the chemical effect, and the smallest power in the contribution of the ES barrier. This is consistent with the fact that the influence of the Schwoebel effect is greater when the kinetics of adatom attachment/detachment is slower (i.e., with decreasing $\ok$) and steps are impermeable ($\okp=0$). By contrast, faster attachment/detachment kinetics (increasing $\ok$) or greater permeability (increasing $\okp$) enhances the influence of the energetic effect (elasticity) at the expense of its kinetic counterparts (Schwoebel barrier and chemical effect).

\subsection{Long-range elasticity}
\label{sub2:range}

\begin{table*}[tb]
\begin{center}
\begin{tabular}{c c c c }
\hline
& $R=1$ & $R=2$ & $R=3$ \\
\hline
\hline
DDE $\quad \displaystyle \frac{\mathrm{Re}(\lambda(k))}{a(k)}$ & $-3\oa$  & $\displaystyle -\frac{(9+\cos(k))3 \oa}{8}$ & $\displaystyle - \frac{(753+113 \cos(k)+16 \cos(2k))3 \oa}{648}$ \\ 
MME $\quad \displaystyle \frac{\mathrm{Re}(\lambda(k))}{a(k)}$ &$\ob$ & $\displaystyle \frac{(3+\cos(k))\ob}{2} $ & $\displaystyle  \frac{(33+17 \cos(k)+4 \cos(2k)) \ob}{18}$ \\  
\hline 
\end{tabular}
\end{center}
\caption{Dependence of the dispersion curve on the range $R$ of elastic interactions considered for steps stabilized solely by elastic dipole-dipole interactions and destabilized solely by their monopole-monopole counterparts, with $S=1$, $\chi =0$, $\okp=0$, and $\onu=0$. DDE refers to dipole-dipole interactions, MME denotes monopole-monopole interactions, and the common factor to the DDE and MME dispersion relations is given $a(k):= \big(16 \ok \Theta \sin^4(k/2) \big) / \big(2+\ok \big)$.}
\label{tab:range}
\end{table*}
Recall that the elastic contribution $\mathfrak{f}_n$ to the configurational force acting on the $n$th step results from the interaction of its elastic field with those generated by the remaining steps on the vicinal surface. However, in the literature on step bunching in the presence of elastic interactions \citep{Tersoff1995,Pierre-Louis2003}, as in Section~\ref{sub2:mech}, stability with respect to bunching is examined under the assumption of nearest-neighbor interactions, whereby the only elastic interactions that are accounted for are those between each step and its immediately adjacent steps. Our objective in this section is to quantify the consequences of this approximation on the stability predictions, both for the dipole-dipole and monopole-monopole interactions.

\begin{figure*}[tb]
\begin{center}
\includegraphics[width=1 \textwidth]{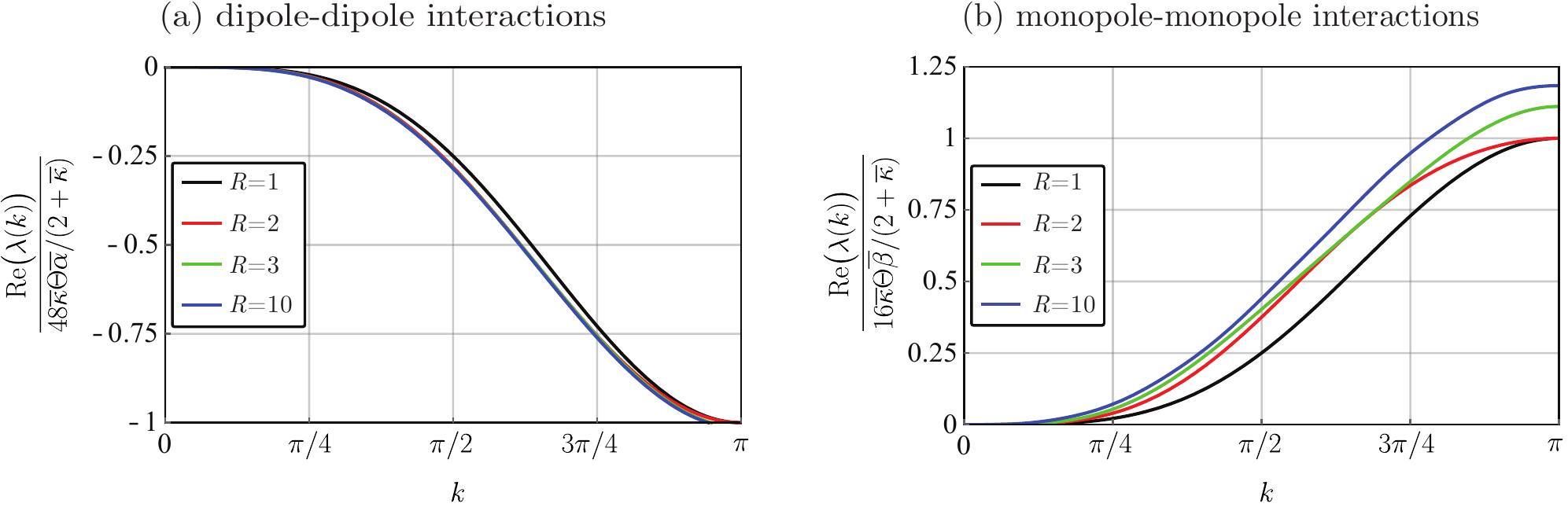}
\caption{\label{fig:range} Dependence of the dispersion curves on the range $R$ of the elastic interactions between steps: (a) dipole-dipole and (b) monopole-monopole interactions. Adatom attachment/detachment is assumed symmetric ($S=1$), steps are taken to be impermeable ($\okp=0$), the chemical effect is formally disabled ($\chi=0$), and sublimation is neglected ($\onu=0$).}
\end{center}
\end{figure*}

For this purpose, we use the linear-stability framework developed in Section~\ref{sub:method} for elastic interactions of arbitrary range $R$ to investigate the dependence of the dispersion curve on $R$ when elasticity is the only mechanism at play. Specifically, the Schwoebel barrier and chemical effects are disabled by setting $S=1$ and $\chi =0$. Moreover, for the sake of obtaining relatively simple analytical expressions for the growth rate of the perturbation, steps are assumed impermeable ($\okp=0$) and evaporation is neglected ($\onu=0$). For both dipole-dipole and monopole-monopole elastic interactions taken independently, the $R$-dependence of the dispersion curve is shown analytically in Table~\ref{tab:range} and graphically on Figure~\ref{fig:range}.

Importantly, the assumptions introduced above do not restrict the generality of our conclusions as the $R$-dependence of the elastic contribution to the stability is not altered when step-step interactions are combined with other mechanisms. Indeed, when the remaining two basic mechanisms are included or sublimation is allowed, the ratio of the contributions to the growth rate of the perturbation of wavelength ${2\pi}/{k}$ of elastic interactions of ranges $R>1$ and $R=1$ remains the same as its counterpart in Table~\ref{tab:range} and Fig.~\ref{fig:range}. For instance, in the presence of evaporation, the growth rate for dipole-dipole interactions with $R=1$ reads
\begin{equation}
\mathrm{Re}(\lambda^{R=1}(k))=\frac{24\ok\Theta\oa\sn\sin^4(k/2)(\ok\cos(k)-\ok\cosh(\sn)-\sn\sinh(\sn)}{2\ok \cosh(\sn)+(\ok^2+\onu)\sinh(\sn)}
\end{equation}
and is modified for second-nearest-neighbor interactions in the proportion
\begin{equation}
\frac{\mathrm{Re}(\lambda^{R=2}(k))}{\mathrm{Re}(\lambda^{R=1}(k))}=\frac{9+\cos(k)}{8}.
\end{equation}
This ratio is the same as the one on display in Table~\ref{tab:range}. The same range dependency in the elastic part of the growth rate is observed when step permeability, the Ehrlich--Schwoebel barrier, and the chemical effect are incorporated in the stability analysis.

As can be seen in Fig.~\ref{fig:range}(a), elastic dipole-dipole interactions with range $R\geq2$ deviate little from their nearest-neighbor counterparts; indeed, convergence of the dispersion curves is reached within less than $1 \%$ at $R=3$. By contrast, the deviation is of the order of $20\%$ in the case of monopole-monopole interactions, as seen in Fig.~\ref{fig:range}(b).

In conclusion, while for the analysis developed here, with the various parameters known only in order of magnitude, the assumption of nearest-neighbor interactions is largely acceptable, the modifications induced by long-range elasticity need to be accounted for if one is to predict accurately the onset of bunching during heteroepitaxy, where monopole-monopole interactions between steps are present.

\section{Discussion}
\label{sec:conclusion}

In Part~I of this investigation we have considered the onset of the bunching instability in the setting of the quasistatic approximation. With few exceptions, this approximation underlies the literature on the stability of vicinal surfaces in the step-flow regime. It serves as the basis of theoretical developments, often without physical justification; in the few works that discuss its validity, it is thought to hold when deposition or evaporation is slow. In the sequel of this two-part article, we will show that it is not the case. Even when the adsorption and desorption rates are small, the terms that are omitted in the quasistatic approximation affect the stability of steps with respect to bunching in significant ways. In Part~I, we therefore take a different viewpoint. By allowing for an almost analytical treatment of the linear-stability problem, the quasistatic approximation is a provisional mathematical simplification that affords us insight into the physics behind step bunching, regardless of its validity. This insight turns out to be useful in interpreting the results of the linear-stability analysis of step flow beyond the quasistatic approximation, as presented in Part~II.

We present here an extension of the BCF model, one that is consistent with the laws of thermodynamics. What this consistency reveals is that the configurational force that drives step motion has an extra contribution, above that which results from the interactions between the elastic fields generated by steps on the vicinal surface. This contribution, which is unaccounted for in most theoretical studies of step instabilities, takes the form of the jump at each step of the grand canonical potential associated with the adatoms on its adjacent terraces. The vanishing of this jump is one of the two conditions of chemical equilibrium, the other being the continuity of the adatom chemical potential. Since steps migrate as a result of the out-of-equilibrium chemical process of adatom attachment and detachment, it is only natural that the jump of the adatom grand canonical potential should appear in the expression of the driving force at steps, and by way of consequence in the expression of the step chemical potential. When adatoms form an ideal lattice gas, this jump is proportional to that of the adatom density. Its presence in the boundary conditions at steps has interesting consequences on their stability against bunching, since it alters the kinetics of adatom attachment and detachment by coupling the diffusion fields on adjacent terraces. We refer to it as the chemical effect.

In our analysis of the stability of steps against bunching, we have distinguished between \emph{mechanisms}, \emph{regimes}, and \emph{parameters}. The mechanisms are three: the \emph{Ehrlich--Schwoebel barrier}, the \emph{elastic step-step interactions}, and the \emph{chemical effect}. In the deposition regime, we show that a direct ES barrier is stabilizing and its inverse destabilizing, the chemical effect is destabilizing, and the elastic dipole-dipole interactions between steps are stabilizing in contrast to their monopole-monopole counterparts which are destabilizing. In the evaporation regime, the influence of the ES barrier on the onset of the bunching instability is opposite, and so is that of the chemical effect, whereas the impact of the elastic step-step interactions on stability remains the same. 

The dimensionless parameters that enter the moving-boundary problem fall in two categories. We call the \emph{deposition} 
and \emph{evaporation} rates \emph{operating parameters}, and refer to the \emph{equilibrium adatom coverage}, the \emph{attachment/detachment coefficient}, and the \emph{step permeability} as \emph{material parameters}. Our stability analysis reveals that the contributions of the ES barrier and chemical effect to the growth rate of linear perturbations vary (linearly) with the operating parameters, whereas the contribution of the elastic interactions between steps does not depend on them. We therefore refer to the first two mechanisms as \emph{kinetic} and label the third mechanism \emph{energetic}. 

The material parameters are neither stabilizing nor destabilizing per se. Instead, they alter the balance between the aforementioned stabilizing and destabilizing mechanisms, when these are simultaneously present (as expected in real experiments). For example, when steps are near transparent to the hopping of adatoms between terraces, elasticity becomes predominant at the expense of the ES barrier and the chemical effect. Moreover, we show that the contribution of the chemical effect to the dispersion relation scales quadratically with the equilibrium adatom coverage, whereas the contributions associated with the remaining two mechanisms are linear in the adatom coverage $\Theta$. The chemical effect is therefore expected to be important on material surfaces with high coverage, such as GaAs(001) and Si(111)-1\texttimes1.

The above results are contingent on our conjecture that the stability of the step-pairing and long-wavelength perturbations implies that of perturbations of all wave numbers.

Finally, we have investigated the validity of the nearest-neighbor approximation for elastic interactions and have concluded that it holds for dipole-dipole interactions but not for their monopole-monopole counterparts. This makes it a reasonable assumption in predictive models for homoepitaxy, where only dipole-dipole interactions are present, but not in a quantitative theory of heteroepitaxy, where monopole-monopole interactions need to be accounted for. \\

\section*{Acknowledgment}

This work is supported by the ``IDI 2015'' project funded by the IDEX Paris-Saclay under grant ANR-11-IDEX-0003-02.

\bibliography{bib_JMPS1and2}

\end{document}